
\input lanlmac

\newcount\figno
\figno=0
\def\fig#1#2#3{
\par\begingroup\parindent=0pt\leftskip=1cm\rightskip=1cm\parindent=0pt
\baselineskip=11pt
\global\advance\figno by 1
\midinsert
\epsfxsize=#3
\centerline{\epsfbox{#2}}
\vskip 12pt
\centerline{{\bf Fig. \the\figno:} #1}\par
\endinsert\endgroup\par
}
\def\figlabel#1{\xdef#1{\the\figno}}


\def\th{\theta}

\def\cob{\delta}

\def\r{{\bf r}}

\def\Tr{{\rm Tr}}

\def\hf{{1\over 2}}

\def\rR{{\rm R}}
\def\d{{\rm d}}

\def\R{{\bf R}}
\def\o{\over}

\def\si{\sigma}

\def\b#1{\overline{#1}}

\def\bra{\langle}
\def\ket{\rangle}
\def\lf{\left}
\def\ri{\right}

\def\al{\alpha}

\def\vphi{\varphi}

\def\Om{\Omega}
\def\l{\ell}
\def\dag{\dagger}
\def\rt#1{\sqrt{#1}}

\def\st{\star}

\def\sitarel#1#2{\mathrel{\mathop{\kern0pt #1}\limits_{#2}}}

\def\bz{\b{z}}


\def\Tr{{\rm Tr}}
\def\s{\sigma}
\def\d{\delta}
\def\eqs{\eqalign}
\def\'{\prime}

\def\to{\rightarrow}
\def\l{\left}
\def\r{\right}
\def\({\left(}
\def\){\right)}
\def\[{\left[}
\def\]{\right]}
\def\a{a^{\dag}}
\def\A{A^{\dag}}
\def\rL{{\rm L}}
\def\rR{{\rm R}}

\def\bz{\bar{z}}
\def\kket{\ket\ket}
\def\bbra{\bra\bra}
\def\stvac{|\Omega_{\rm str}\ket\ket}
\def\stbac{\bra\bra\Omega_{\rm str}|}

\def\np#1#2#3{{ Nucl. Phys.} {\bf B#1} (#2) #3}

\def\plo#1#2#3{{ Phys. Lett.} {\bf #1B} (#2) #3}

\def\prl#1#2#3{{ Phys. Rev. Lett.} {\bf #1} (#2) #3}


\lref\Witten{
E. Witten,
``Noncommutative Geometry And String Field Theory,''
Nucl.\ Phys.\  {\bf B268} (1986) 253.
}
\lref\LPP{
A. LeClair, M. E. Peskin and C. R. Preitschopf,
``String Field Theory on the Conformal Plane. 1. Kinematical Principles,''
Nucl. Phys. {\bf B317} (1989) 411\semi
``String Field Theory on the Conformal Plane. 2. Generalized Gluing,''
Nucl. Phys. {\bf B317} (1989) 464.
}
\lref\HorowitzLRS{
G. T. Horowitz, J. Lykken, R. Rohm and A. Strominger,
``Purely Cubic Action for String Field Theory,''
\prl{57}{1986}{283}.
}
\lref\Samuel{
S.~Samuel,
``The Physical and Ghost Vertices in Witten's String Field Theory,''
\plo{181}{1986}{255}.
}
\lref\GrossJ{
D. J. Gross and A. Jevicki,
``Operator Formulation of Interacting String Field Theory,''
\np{283}{1987}{1}\semi
``Operator Formulation of Interacting String Field Theory (II),''
\np{287}{1987}{225}.
}
\lref\CST{
E.~Cremmer, A.~Schwimmer and C.~Thorn,
``The Vertex Function in Witten's Formulation of String Field Theory,''
\plo{179}{1986}{57}.
}
\lref\ItohWM{
K.~Itoh, K.~Ogawa and K.~Suehiro,
``BRS Invariance of Witten's Type Vertex,''
Nucl.\ Phys.\  {\bf B289} (1987) 127.
}
\lref\OhtaWN{
N.~Ohta,
``Covariant Interacting String Field Theory in the Fock Space Representation,''
Phys.\ Rev.\ D {\bf 34} (1986) 3785 
[Erratum-ibid.\ D {\bf 35} (1986) 2627].
}

\lref\StromingerBN{
A.~Strominger,
``Point Splitting Regularization Of Classical String Field Theory,''
Phys.\ Lett.\  {\bf B187} (1987) 295.
}
\lref\HorowitzYZ{
G.~T.~Horowitz and A.~Strominger,
``Translations As Inner Derivations And Associativity Anomalies In Open
String Field Theory,''
Phys.\ Lett.\  {\bf B185} (1987) 45.
}
\lref\GMS{
R. Gopakumar, S. Minwalla and A. Strominger,
``Noncommutative Solitons,''
JHEP {\bf 0005} (2000) 020, {\tt hep-th/0003160}.
}
\lref\DasguptaMR{
K. Dasgupta, S. Mukhi and G. Rajesh,
``Noncommutative Tachyons,''
JHEP {\bf 0006} (2000) 022, {\tt hep-th/0005006}.
}
\lref\HKLM{
J. A. Harvey, P. Kraus, F. Larsen and E. J. Martinec,
``D-branes and Strings as Non-commutative Solitons,''
JHEP {\bf 0007} (2000) 042, {\tt hep-th/0005031}.
}
\lref\Harvey{
J. A. Harvey,
``Komaba Lectures on Noncommutative Solitons and D-Branes,''
{\tt hep-th/0102076}.
}
\lref\HKL{
J. A. Harvey, P. Kraus and F. Larsen,
``Exact Noncommutative Solitons,''
JHEP {\bf 0012} (2000) 024, {\tt hep-th/0010060}.
}
\lref\Furuuchi{
K. Furuuchi,
``Topological Charge of U(1) Instantons,''
{\tt hep-th/0010006}.
}
\lref\KT{
T. Kawano and T. Takahashi,
``Open String Field Theory on Noncommutative Space,''
Prog. Theor. Phys. {\bf 104} (2000) 459, {\tt hep-th/9912274}\semi
``Open-Closed String Field Theory in the Background B-Field,''
Prog. Theor. Phys. {\bf 104} (2000) 1267,
{\tt hep-th/0005080}.
}
\lref\Sugino{
F. Sugino,
``Witten's Open String Field Theory in Constant B-Field Background,''
JHEP {\bf 0003} (2000) 017, {\tt hep-th/9912254}.
}
\lref\Snbl{
M. Schnabl,
``String Field Theory at Large $B$-field and Noncommutative Geometry,''
JHEP {\bf 0011} (2000) 031, {\tt hep-th/0010034}.
}
\lref\WittenNCG{
E. Witten,
``Noncommutative Tachyons And String Field Theory,''
{\tt hep-th/0006071}.
}
\lref\WittenK{
E. Witten,
``Overview Of K-Theory Applied To Strings,''
Int. J. Mod. Phys. {\bf A16} (2001) 693, {\tt hep-th/0007175}.
}
\lref\WittenDK{
E.~Witten,
``D-branes and K-theory,''
JHEP {\bf 9812} (1998) 019,
{\tt hep-th/9810188}.
}
\lref\RastelliZ{
L. Rastelli and B. Zwiebach,
``Tachyon potentials, star products and universality,''
{\tt hep-th/0006240}.
}
\lref\RSZone{
L. Rastelli, A. Sen and B. Zwiebach,
``String Field Theory Around the Tachyon Vacuum,''
{\tt hep-th/0012251}.
}
\lref\RSZtwo{
L. Rastelli, A. Sen and B. Zwiebach,
``Classical Solutions in String Field Theory Around the Tachyon Vacuum,''
{\tt hep-th/0102112}.
}
\lref\KP{
V. A. Kostelecky and R. Potting,
``Analytical construction of a nonperturbative vacuum
for the open bosonic string,''
Phys. Rev. {\bf D63} (2001) 046007, {\tt hep-th/0008252}.
}
\lref\RSZthree{
L. Rastelli, A. Sen and B. Zwiebach,
``Half-strings, Projectors, and Multiple D-branes
in Vacuum String Field Theory,''
{\tt hep-th/0105058}.
}
\lref\GT{
D. J. Gross and W. Taylor,
``Split string field theory I,''
{\tt hep-th/0105059}.
}
\lref\ChanIE{
H.~Chan and S.~T.~Tsou,
``String Theory Considered As A Local Gauge Theory Of An Extended Object,''
Phys.\ Rev.\ D {\bf 35}, 2474 (1987)\semi
``Yang-Mills Formulaton Of Interacting Strings,''
Phys.\ Rev.\ D {\bf 39}, 555 (1989).
}
\lref\BordesXH{
J.~Bordes, H.~Chan, L.~Nellen and S.~T.~Tsou,
``Half String Oscillator Approach to String Field Theory,''
Nucl.\ Phys.\  {\bf B351} (1991) 441.
}
\lref\AbdurrahmanGU{
A.~Abdurrahman and J.~Bordes,
``The relationship between the comma theory and 
Witten's string field  theory. I,''
Phys.\ Rev.\  {\bf D58} (1998) 086003.
}
\lref\ArefevaEG{
I.~Y.~Arefeva and I.~V.~Volovich,
``Two-dimensional gravity, string field theory and spin glasses,''
Phys.\ Lett.\ B {\bf 255}, 197 (1991).
}
\lref\KOO{
T. Kawano, K. Ohmori and K. Okuyama, 
unpublished.
}

\Title{                                            \vbox{\hbox{KEK-TH-766}
                                                   \hbox{UT-937}
                                                   \hbox{\tt hep-th/0105129}}}
{\vbox{
\centerline{Open String Fields As Matrices}
}}


\centerline{Teruhiko Kawano}
\vskip 1mm
\centerline{\sl Department of Physics, University of Tokyo}
\vskip -0.05in
\centerline{\sl Hongo, Tokyo 113-0033, Japan}
\centerline{\tt kawano@hep-th.phys.s.u-tokyo.ac.jp}

\vskip .1in
\centerline{and}
\vskip .1in

\centerline{Kazumi Okuyama}
\vskip 1mm
\centerline{\sl Theory Group, KEK}
\vskip -0.05in
\centerline{\sl Tsukuba, Ibaraki 305-0801, Japan}
\centerline{\tt kazumi@post.kek.jp}

\vskip 1in
\noindent
We present a new representation of the string vertices of
the cubic open string field theory.
By using this three-string vertex, we attempt to identify 
open string fields as huge-sized matrices by following Witten's idea.
By using these huge matrices, we obtain some results about 
the construction of partial isometries in the algebra of open string fields.

\vskip .2in
\Date{May, 2001}

\vfill
\vfill

\newsec{Introduction}

In the paper \RSZone, Rastelli, Sen and Zwiebach have proposed
the cubic open string field theory \Witten\ around the tachyon vacuum,
where it is supposed that only closed strings exist with no 
open strings. In the subsequent work \RSZtwo, they found
$D$-branes as classical solutions in their proposed open string field theory
by using the projection fields obtained in \KP\ for the matter sector.
The analogous technique where projection operators are used to find 
classical
solutions has also been seen in noncommutative field theories 
\refs{\GMS,\HKL}.
Especially in the latter paper \HKL, this solution generating technique has 
been used to find solutions in noncommutative gauge theory.
This solution generating technique needs the system under consideration
to have gauge invariance. Since the cubic open string field theory
is invariant under huge gauge symmetry, it seems to be very useful
in finding classical solutions analytically.
In fact, the authors of \refs{\Snbl,\HKL} have argued the possibility
that such techniques can be applied to  the cubic string field theory and
the boundary string field theory (See also \KP\ for another approach).

Let us briefly review their discussion about the application of
the solution generating technique to the cubic string field theory.
The variation of the action \Witten\
\eqn\action{
S(A)={1\over g_{\rm s}^2}\Tr\l[\hf A\st QA+{1\over3}A\st A\st A\r],
}
with respect to string field $A$, gives the equation of motion
\eqn\eom{
{\delta S \over\delta A}=QA+A\st A=0.
}
The action \action\ is invariant under the infinitesimal gauge
transformation
\eqn\igtrf{
-i\delta_{\Lambda}A=Q\Lambda+A\st\Lambda-\Lambda\st A,
}
and its finite form can be seen to be
\eqn\gtrf{
A \to A'= U^{\dag}\st QU+U^{\dag}\st A\st U
}
where
\eqn\UE{
U=\exp_{\st}\(-i\Lambda\)=\sum^{\infty}_{n=0}{(-i)^n\over n!}\
\underbrace{\Lambda\st\Lambda\st\cdots\st\Lambda}_{n\ {\rm times}}.
}

Now we turn to the solution generating technique with 
two operators, $U$ and $V$, satisfying
\eqn\sgt{
U \st V=I, \qquad V \st U=I-P,
}
where $P$ is an operator such that $QP=0$. 
Then we consider the field redefinition
\eqn\chA{
A'=V\st QU+V\st A\st U
}
in the equation of motion \eom\ and obtain
\eqn\cheom{
{\delta S \over\delta A}\(A'\)=V\st{\delta S \over\delta A}\(A\)\st U,
}
which shows that
if $A$ is a solution to \eom, as far as $P\not=0$, $A'$ gives a
new solution to it.
The above condition $QP=0$ is required to obtain \cheom, 
otherwise $A'$ would no longer be a solution even if $A$ is a 
solution to the equation of motion \eom. 
The simplest solution of \eom\ is $A=0$.
Therefore, through the field redefinition \chA, we can obtain
another solution
\eqn\VQU{
A_c=V \st QU,
}
for any $U$ and $V$ satisfying \sgt.
The action associated with the solution \VQU\ turns out to be
\eqn\actionA{
S\(A_c\)=-{1\over6g_{\rm s}^2}\Tr\[V\st QU\st V\st QU\st V\st QU\].
}
If $D$-brane solutions could be obtained in this way, we could gain 
a deeper understanding of the tachyon condensation and
the $D$-brane dynamics in the cubic string field theory.


In the cubic string field theory, our present technology does not seem enough 
to find such partial isometries $U$, $V$, let alone the nontrivial classical 
solutions to \eom. In the case of the field theories \refs{\GMS,\HKL}, 
under the Weyl transformation, fields are transformed to operators acting 
on the ordinary Hilbert space. Such operators can also be given as matrices 
in terms of states in the Hilbert space. Therefore, the Moyal product of fields 
is transformed to the usual matrix multiplication, 
which seems to make it easy to find projection operators 
and partial isometries. Thus, to develop analogous methods in the string field 
theory, we would be happy if we could have the formulation where open string 
fields can be treated as matrices. Recalling that the open string product 
$A\st{B}$ of fields $A$, $B$ is defined \Witten\ such that the right half 
of the string of $A$ is identified with the left half of $B$ and summed over all 
the possible configurations to give the third string $A\st{B}$, 
we are very tempted to consider the right half of the string field $A$ 
as the column of some huge matrix and the left of $B$ as the row. 
In fact, Witten have already discussed the above idea in his paper \Witten. 
In this paper, we will try to give an explicit realization of this idea 
by introducing a new representation of string vertices. 
Although we believe that our attempt is successful to some extent, 
there might exist some subtleties in our treatment of the midpoint of string 
vertices, as will be discussed in section 5.  

In section 2, we will present the new string vertices for the matter sector, 
which have a simpler form compared to the usual oscillator representation 
\refs{\Samuel\GrossJ\CST\ItohWM{--}\OhtaWN}. 
The kinetic term (the Virasoro operator) $L_0$ in a flat Minkowski spacetime 
becomes diagonal in terms of the usual oscillators of the string coordinates. 
In this paper, we introduce another set of oscillators. 
These oscillators do not any longer diagonalize the above Virasoro operator, 
but make string vertices very simple. 
Since our attention will be paid mainly to the open string product $\st$, 
the new oscillators will be adopted for our study in this paper. 
In addition, we have not found such simple vertices for the ghost sector. 
Since the open string product does not mix the ghost operators with ones in 
the matter sector, we will be content with the study of the open string 
product only in the matter sector. 

In section 3, we will obtain the mapping from string fields to 
huge-sized matrices by using the above-mentioned three-string vertex. 
In section 4, taking advantage of the matrix representation of string fields, 
we will construct some projectors and partial isometries, and 
discuss the relation of these operators to $D$-branes. 
Most of section 5 is devoted to the discussion about potential 
subtleties in this work.

\newsec{A New Representation of String Vertices}

In this section, we shall introduce a new representation of the 
three-string
vertex, which is given in terms of unusual oscillator modes of string
coordinates. Although the BRS charge in a flat Minkowski spacetime is not
diagonalized by the new oscillators, the three-string vertex becomes simple
in terms of them. The simplicity of our string vertices turns out to
give us more insights into the structure of the open string product.

\subsec{Preliminary}

To discuss the new representation of string vertices,
we introduce unconventional oscillators in the
Hilbert space of a single open string.
Let us consider an open string parametrized by
$\si\in[0,\pi]$ with the Neumann boundary condition at $\si=0,\pi$.
The ordinary mode expansion of the coordinate $X^i(\si)$ and the momentum
$P_j(\si)$ is given by
\eqn\modeXP{\eqalign{
X^i(\si)&=x^i+i\rt{2\al'}\sum_{n\not=0}{1\o n}\al_n^i\cos n\si,\cr
P_j(\si)&={1\o\pi}\lf(p_j+{1\o\rt{2\al'}}
\sum_{n\not=0}\eta_{jk}\al^k_n\cos n\si\ri).
}}
The canonical commutation relation is
\eqn\XPcom{
[X^i(\si),P_j(\si')]=i\d^i_j\cob(\si,\si'),
}
where $\cob(\si,\si')$ is the $\cob$-function satisfying the
Neumann boundary condition:
\eqn\cobincos{
\cob(\si,\si')={1\o\pi}\sum_{n=-\infty}^{\infty}\cos n\si\cos n\si'.
}
From \XPcom, the commutation relations of the modes are found to be
\eqn\comaln{
[x^i,p_j]=i\d^i_j,\quad [\al^i_n,\al^j_m]=n\eta^{ij}\cob_{n+m,0}.
}
The $SL(2,\R)$ vacuum $|0\ket$ is defined by
\eqn\SLvacdef{
p_j|0\ket=0,\quad \al^i_n|0\ket=0~~(n\geq 1).
}

Now, let us introduce a new annihilation operator $a(\si)$ by
\eqn\defasi{
a^i(\si)={1\o\rt{2}}\lf(\rt{2\pi\al't}P^i(\si)-i{X^i(\si)\o\rt{2\pi\al't}}\ri),
}
where $t$ is an arbitrary parameter\foot{For the zero-mode part,
the parameter $b$ in \RSZtwo\ corresponds to $4\al't$.}.
Then, the commutator of $a^i(\si)$ and its hermitian conjugate
$a^{\dag}_j(\si)$ becomes
\eqn\comaadag{
[a^i(\si),a^{\dag}_j(\si')]=\d^i_j\cob(\si,\si').
}
Using $a^i(\si)$, we define the ``$a$-vacuum''  $|\Om\ket$  by
\eqn\avacdef{
a^i(\si)|\Om\ket=0,\qquad {\rm for\ }\forall\si\in[0,\pi].
}

It is well-known that the $SL(2,\R)$ vacuum $|0\ket$ and the $a$-vacuum
$|\Om\ket$ are related by the Bogoliubov transformation.
A set of functions $\{\varphi_n(\si)\}$ given on the interval $[0,\pi]$ by
\eqn\rephidef{
\varphi_0(\si)={1\o\rt{\pi}},\quad \varphi_n(\si)=\rt{2\o\pi}\cos n\si 
}
provide all the functions on $[0, \pi]$ with the Neumann boundary condition
at $\s=0,\pi$ with a complete basis, in terms of which any of those functions
can therefore be expanded.
They are orthogonal on $[0,\pi]$:
\eqn\intpiphi{
\int_0^{\pi}d\si\,\varphi_n(\si)\,\varphi_m(\si)=\cob_{n,m},
}
and are summed over to give the $\cob$-function
\eqn\phitocob{
\cob(\si,\si')=\sum_{n=0}^{\infty}\,\varphi_n(\si)\,\varphi_n(\si').
}
The annihilation operator
$a^i(\si)$ can thus be expanded by $\varphi_n$ as
\eqn\expaphi{
a^i(\si)=\sum_{n=0}^{\infty}a_n^i\,\varphi_n(\si).
}
From \comaadag, \phitocob\ and \expaphi,
we obtain the commutation relation
\eqn\comman{
[a_n^i,a_{mj}^{\dag}]=\d^i_j\cob_{n,m}.
}
Using \modeXP\ and \defasi, we can see that
$a^i_n$ and $\al^i_n$ are related by
\eqn\atoaln{\eqalign{
a^i_0&=\rt{\al't}p^i-i{x^i\o2\rt{\al't}}, \cr
a^i_n&=\hf\lf(\rt{t}+{1\o n\rt{t}}\ri)\al^i_n
+\hf\lf(\rt{t}-{1\o n\rt{t}}\ri)\al^i_{-n}.
}}
By introducing the oscillators

\eqn\defbn{
b^i_n={1\o\rt{n}}\al^i_n,\quad {b_n^i}^{\dag}={1\o\rt{n}}\al^i_{-n}~~(n\geq 
1)
}
satisfying $[b^i_n,b_{mj}^{\dag}]=\d^i_j\cob_{n,m}$,
the relation \atoaln\  can be rewritten as
\eqn\antobn{
a^i_n=b^i_n\cosh\th_n+{b^i_n}^{\dag}\sinh\th_n
}
where $\th_n$ is given by
\eqn\defthn{
e^{\th_n}=\rt{tn}.
}
Therefore, $a^i_n$ and $b^i_n$ are related by
the Bogoliubov transformation
\eqn\abogob{
a^i_n=Ub^i_nU^{-1},
}
where $U$ is a unitary operator
\eqn\defUbogtrf{
U=\exp\lf(\hf\sum_{n=1}^{\infty}\th_n(b_n^2-b_n^{\dag 2})\ri)
=\exp\lf({1\o4}\sum_{n=1}^{\infty}{\log tn\o n}(\al_n^2-\al_{-n}^2)\ri).
}
For the zero-mode part, the zero momentum eigenstate
$|p=0\ket$ and $a$-vacuum are  related by
\eqn\pzeroazero{
|p=0\ket=e^{-\hf a_0^{\dag2}}|\Om\ket.
}
Note that the transformation in the zero-mode part is not unitary.
After all, the relation between $|\Om\ket$ and $|0\ket$ is found to be
\eqn\mapomtozero{
|\Om\ket=e^{\hf a_0^{\dag2}}U|0\ket.
}

\subsec{The String Vertices}

A string vertex is determined up to an overall normalization
by an overlap condition. For the midpoint interaction as in Witten's
open string field theory \Witten, the overlap condition is given by
\eqn\midpoint{
X_r^i(\s)-X_{r-1}^i(\pi-\s)=0, \qquad P_r^i(\s)+P_{r-1}^i(\pi-\s)=0\
}
for $\s\in$ L, where L denotes the interval $[0, \pi/2]$ and
R will be used to denote $[\pi/2, \pi]$. From \defasi, we find the overlap
condition
\eqn\overlap{
a^i_r(\s)=-{a^i_{r-1}}^{\dag}(\pi-\s)\equiv-r({a^i_{r-1}}^{\dag})(\s)
\quad {\rm for}\ \s\in \rL.
}
For the notational simplicity, we henceforth omit vector indices $i, j$ of
the Lorentz group.

The overlap condition \overlap\ is solved by the $N$-string vertex
\eqn\vertex{\eqs{
|v^{m}_N&(1,\cdots,N)\ket=e^{-V_N}|\Om\ket_{1\cdots N};
\cr
V_N&=\sum^{N}_{r=1}\int^{\pi\o2}_{0} d\s\, \a_r(\s)\a_{r-1}(\pi-\s)
\cr
&=\hf\sum^N_{r,s=1}\int^{\pi}_{0} d\s d\s'\,
\a_r(\s)N^{rs}(\s,\s')\a_{s}(\s').
\cr}
}
Here, the definition of the function $N^{rs}(\s,\s')$ is
\eqn\neuman{
N^{rs}(\s,\s')=\[\d_{r-1,s}\theta_{\rL}(\s)
+\d_{r+1,s}\theta_{\rR}(\s)\]\d(\pi-\s,\s'),
}
where $\theta_{\rL}(\s)$ and $\theta_{\rR}(\s)$ are step functions with
their supports $[0, \pi/2]$ and $[\pi/2, \pi]$, respectively.
More precisely, the Fourier expansion of the step functions
$\theta_{{\rL},{\rR}}(\s)$ is given by
\eqn\fttheta{
\theta_{{\rL,\rR}}(\s)=\hf\pm{2\over\pi}\sum^{\infty}_{n=1}{1\over n}
\sin\l({\pi\over2}n\r)\cos\l(n\s\r).
}
Incidentally, by using the notation in \overlap, we can express $V_N$ as
\eqn\svertex{
V_N=\sum^{N}_{r=1}\int_{\rL}\a_rr(\a_{r-1}),
}
where
$\int_{\rL}$ means $\int^{\pi\o2}_0d\s$.
The vertices \svertex\ thus have very simple forms compared to 
the usual oscillator representation \refs{\Samuel\GrossJ\CST\ItohWM{--}\OhtaWN}. 

Before going further, let us make a small digression on a few computations
to show how our vertices
facilitate the computations which have been supposed to be
very tedious. As the first computation, we pick up the sewing procedure of
two string vertices by using the reflector $\bra R(I,J)|$
to obtain one string vertex. Indeed, we shall prove the equation
\eqn\rvv{
\bra R(K,L)||v^m_{M+1}(1,\cdots,M,K)\ket
|v^m_{N+1}(L,M+1,\cdots,M+N)\ket=|v^m_{M+N}(1,\cdots,M+N)\ket.
}
The reflector $\bra R(I,J)|$ satisfies the defining condition
$\bra R(1,2)|a_r(\s)=-\bra R(1,2)|r(\a_{r-1})(\s)$ and is found to be
\eqn\reflector{\eqs{
\bra R(1,2)|
&=
\ _{12}\bra \Om|\exp\[-\sum^{2}_{r=1}
\int^{\pi\o2}_0d\s\, a_r(\s)a_{r-1}(\pi-\s)\]
\cr
&=\ _{12}\bra \Om|\exp\[-\int^{\pi}_0d\s\, a_1(\s)a_{2}(\pi-\s)\].
\cr}
}
By virtue of the formula
$\bra \Om|e^{\lambda\cdot a}e^{\a\cdot\mu}|\Om\ket
=e^{\lambda\cdot\mu}$,
where $\lambda\cdot\mu=\int^{\pi}_0d\s\lambda(\s)\mu(\s)$,
we can see that
\eqn\proofvvr{\eqs{
&\bra R(K,L)||v^m_{M+1}(1,\cdots,M,K)\ket |v^m_{N+1}(L,M+1,\cdots,M+N)\ket
\cr
&\sim \ _{KL}\bra \Om|e^{-a_K\cdot r(a_L)}
e^{-\int_\rL[\a_Kr(\a_M)+\a_1r(\a_K)]}
e^{-\int_\rL[\a_{M+1}r(\a_L)+\a_Lr(\a_{M+N})]}|\Om\ket_{KL}
\cr
&=\ _L\bra \Om|e^{\int_{\rL}[r(\a_M)r(a_L)+\a_1a_L]}
e^{-\int_{\rL}[\a_{M+1}r(\a_L)+\a_Lr(\a_{M+N})]}|\Om\ket_L
\cr
&=e^{-\int_L[\a_{M+1}r(\a_M)+\a_1r(\a_{M+N})]},
\cr}
}
which proves \rvv.

The next computation is concerned with the `identity' field
\eqn\iden{
\bra I(K)|=\bra \Om|\exp\[-\hf a_K\cdot r(a_K)\]
}
satisfying the overlap condition
\eqn\condiden{
\bra I(K)|\a_K(\s)=-\bra I(K)|r(a_K)(\s)
}
for the $K$-th string.
The use of the identity field allows us to obtain 
\eqn\ivv{
\bra I(N)||v^{m}_N(1,\cdots,N-1,N)\ket=|v^{m}_{N-1}(1,\cdots,N-1)\ket.
}
In order to prove this identity, the coherent state
\eqn\coherent{\eqs{
&a(\s)|z\ket=z(\s)|z\ket
\cr
&|z\ket=\exp\[\int^{\pi}_0d\s\a(\s)z(\s)\]|\Om\ket 
\cr}
}
will be useful.
Here $z(\s)$ is expanded as $z(\s)=\sum^{\infty}_{r=0}z_n\varphi_n(\s)$.
Also, its hermitian conjugate $\bra{z}|$ satisfies 
\eqn\bracoherent{
\bra z|\a(\s)=\bra z|\bz(\s).
}
Note that $\bra z|\not=\ _1\bra r(z)|\equiv\bra R(1,2)|z\ket_2$.
Then, we can verify
\eqn\completebasis{
\int \[dzd\bar{z}\]\, |z\ket e^{-\int^\pi_0d\s|z(\s)|^2}\bra z| =1
}
where the measure is given by $[dzd\bar{z}]=\prod_{n=0}^{\infty}dx_ndy_n$
if $z_n=x_n+iy_n$.

By making use of the coherent state, we can prove the formula
\eqn\aMa{\eqs{
&\exp\[\hf a\cdot M\cdot a+\lambda\cdot a\]
\exp\[\hf \a\cdot N\cdot\a +\mu\cdot\a\]|\Om\ket
\cr
&=\sqrt{\det\[1-MN\]}
\exp\[-\hf{\xi}\cdot{\cal M}^{-1}\cdot{\xi}\]
\cr
&\quad\exp\[(\lambda\cdot N+\mu)\cdot(1-MN)^{-1}\cdot\a
+\hf\a\cdot N(1-MN)^{-1}\cdot\a\]|\Om\ket,
\cr}
}
where
\eqn\matrixM{\eqs{
&{\cal M}=\(\matrix{M(\s,\s')&-\d(\s,\s')\cr-\d(\s,\s')&N(\s,\s')}\),
\cr
&{\cal M}^{-1}=-\(\matrix{N(1-MN)^{-1}(\s,\s')&(1-NM)^{-1}(\s,\s')
\cr(1-MN)^{-1}(\s,\s')&M(1-NM)^{-1}(\s,\s')}\),
\cr}
}
and $\xi(\s)=(\lambda(\s), \mu(\s))$.

The formula \aMa\ helps us to calculate
\eqn\proofivv{\eqs{
&\bra I(N)||v^{m}_N(1,\cdots,N-1,N)\ket
\cr
&\sim\ _N\bra \Om|e^{-\hf a_N\cdot r(a_N)}
e^{-\int_{\rL}[\a_N r(\a_{N-1})+\a_1r(\a_N)]}|\Om\ket_N
\cr
&=e^{-\int_{\rL}\a_1r(\a_{N-1})}.
\cr}
}
Thus, it completes the proof of the identity \ivv. 

We conclude this section with the definition of the open string
product. Given two string fields $A$ and $B$, we define
the open string product $A\st{B}$ as
\eqn\starproduct{
|A\st B\ket_1= \ _3\bra r(A)|_2\bra r(B)||v^m_3(1,2,3)\ket,
}
recalling that $_1\bra r(A)|=\bra R(1,2)|A\ket_2$.
For two coherent states $|z\ket$ and $|w\ket$, a small computation
shows that the product $|z\st{w}\ket$ is given by
\eqn\zstarw{
|z\st{w}\ket=e^{-\int_\rL r(z)\cdot{w}}
e^{\int_\rL\(z\a_1+r(w)r(\a_1)\)}|\Omega\ket.
}
If the notation $|z_\rL, z_\rR\ket$ denotes $|z\ket$ with
the functions $z_{\rL,\rR}(\s)=\theta_{\rL,\rR}(\s)z(\s)$,
the above equation \zstarw\ has a simple form
\eqn\zlstarwr{
|z_\rL, z_\rR\ket\st|w_\rL, w_\rR\ket
=e^{-\int_\rL r(z)\cdot{w}}|z_\rL, w_\rR\ket.
}



\newsec{String Field Algebra}

\subsec{String Field Oscillators}
In this section, we construct the Fock space in which all the states
are obtained by acting on the Fock vacuum $\stvac$ with string fields
$A(\si)$ and $A^{\dag}(\si)$. 
These string fields play the same role as the annihilation and creation 
operators, and satisfy the commutation relation
\eqn\commAAd{
[A(\s),A^{\dag}(\s')]_{\st}=\cob(\s,\s')I,\qquad
{\rm for~} \s,\s'\in\rR, 
}
where $[\ ,\ ]_\st$ denotes the commutator with the open string
product:
\eqn\starcommutation{
\[A, B\]_\st = A \st B-B \st A,
}
and $I$ denotes the identity field.
We henceforth call the operators $A(\s)$, $A^{\dag}(\s)$
the ``string field oscillators''.

To seek the string field oscillators,
let us begin with the open string product of fields $I(z)$
given by
\eqn\zidentity{
|I(z)\ket=\exp\[\int^\pi_0 d\s z(\s)\a(\s)\]|I\ket.
}
Making use of \aMa, we find that
\eqn\IzIw{
I(z)\st I(w)=e^{-\int_\rL w\cdot{r(z)}} I(z+w).
}
Taking the derivatives with respect to $z(\s)$ and $w(\s)$ in \IzIw,
we obtain
\eqn\aIaIcomm{
\[ \a(\s)\,I,\ \a(\s')\,I\, \]_\st
=\(\theta_\rL(\s)-\theta_\rR(\s)\)\d(\s,\pi-\s')\,I.
}
Since the L.H.S. of \aIaIcomm\ is antisymmetric under the exchange of
$\s$ and $\s'$, we can see that the R.H.S. is  equal to
$-\(\theta_\rL(\s')-\theta_\rR(\s')\)\d(\s,\pi-\s')\,I$.
Therefore, we have
\eqn\aLaLcomm{\eqs{
&\[ \theta_{\rR}(\s)\a(\s)\,I,\ \theta_{\rR}(\s')\a(\s')\,I\, \]_\st=0,
\cr
&\[ \theta_{\rR}(\s)\a(\s)\,I,\ -\theta_{\rR}(\s')\a(\pi-\s')\,I\, \]_\st
=\theta_{\rR}(\s)\d(\s,\s')\theta_{\rR}(\s')\,I,
\cr
&\[ \theta_{\rR}(\s)\a(\pi-\s)\,I,\ \theta_{\rR}(\s')\a(\pi-\s')\,I\, 
\]_\st
=0.
}}
The string field oscillators can thus be identified as
\eqn\stringosci{
A(\s)=\theta_{\rR}(\s)\a(\s)\,I, \qquad 
\A(\s)=-\theta_{\rR}(\s)r(\a)(\s)\,I.
}
Here we notice that the identity $I$ maps the
right part of the worldsheet oscillator $\a(\s)$ into
the string field annihilation operator $A(\s)$ and the left part
into the creation operator $A^{\dag}(\s)$.
Note that the form of $A(\s)$ is somehow
reminiscent of the nilpotent
string field $Q_{\rL}I$
\HorowitzLRS, where $Q$ is the BRS charge.

After the substitution 
$z(\s)\to z(\s)\theta_\rR(\s)$ and $w(\s)\to w(\s)\theta_\rR(\s)$ 
in \IzIw, the derivatives at $z(\s)=w(\s)=0$ with respect to 
both of $z(\s)$ and $w(\s)$ give 
\eqn\AstAtoaaI{\eqalign{
&A(\si)\st A(\s')=\theta_\rR(\s)\a(\s)\,\theta_\rR(\s')\a(\s')\,I,
\cr
&\A(\si)\st \A(\s')=\theta_{\rR}(\s)r(\a)(\s)\, 
\theta_{\rR}(\s')r(\a)(\s')\,I.
}}
The recursive use of \AstAtoaaI\ provides
\eqn\expstA{\eqalign{
&\exp_{\st}\lf({\int^{\pi}_0d\s\, w(\s)A^{\dag}(\s)}\ri)
=e^{-\int_{\rR} w\cdot r(\a)}\,I,
\cr
&\exp_{\st}\lf({\int^{\pi}_0d\s\, \bz(\s)A(\s)}\ri)
=e^{\int_{\rR} \bz\cdot\a}\,I.
}}
Here we have defined the exponential of string field $M$ by
\eqn\defexpstar{
\exp_{\st}(M)\equiv\sum_{n=0}^{\infty}\,{1\o n!}\,
\underbrace{M\st M\st\cdots \st M}_{n\ {\rm times}}.
}

Taking a step further, we look for the Fock vacuum $\stvac$ of
the string field oscillators;
\eqn\fockvac{
A(\s)\stvac=0.
}
To this end, it is appropriate to consider the open string product
\eqn\wIzIzw{\eqs{
|{w}_\rL, {w}_\rR\ket\st|{I(z)}\ket=e^{-\int_\rR w\cdot{r(z)}}
|{w}_\rL, {w}_\rR+{z}_\rR\ket,
\cr
|{I(z)}\ket\st|{w}_\rL, {w}_\rR\ket=e^{-\int_\rR z\cdot{r(w)}}
|{z}_\rL+{w}_\rL, {w}_\rR\ket.
}}
In \wIzIzw, replacing $z(\s)$ with $z(\pi-\s)\theta_{\rL}(\s)$ in the
first equation and with $z(\s)\theta_{\rR}(\s)$ in the second,
we take the first derivative with respect to $z(\s)$ at $z(\s)=w(\s)=0$. 
Then we find that
\eqn\stringvac{
|\Omega\ket\st\A(\s)=0, \quad A(\s)\st|\Omega\ket=0,
}
which, together with \fockvac, indicates that
\eqn\strketbra{
|\Omega\ket \sim \stvac\stbac, 
}
recalling here that $|\Om\ket$ is the $a$-vacuum in \avacdef.
Furthermore, combining \expstA\ and \wIzIzw, we find 
\eqn\LRgenerator{\eqs{
|w_\rL,w_\rR\ket\st\({e_\st}^{\int_\rR z\cdot{A}}\)=|w_\rL,w_\rR+z_\rR\ket,
\cr
\({e_\st}^{-\int_\rR 
r(z)\cdot{\A}}\)\st|w_\rL,w_\rR\ket=|z_\rL+w_\rL,w_\rR\ket.
}}
From these equations \LRgenerator, we understand that
the coherent field $|z\ket$ can be written as
\eqn\strG{\eqs{
|z_\rL, z_\rR\ket&=\({e_\st}^{-\int_\rR r(z)\cdot{\A}}\)\st|\Omega\ket
\st\({e_\st}^{\int_\rR z\cdot{A}}\)
\cr
&=\({e_\st}^{-\int_\rR r(z)\cdot{\A}}\)\stvac\stbac
\({e_\st}^{\int_\rR z\cdot{A}}\).
}}
Note that the open string product $\st$ in the first line is replaced 
by the usual matrix multiplication in the second line in \strG. 

As a simple consistency check, we can verify that the equation \zlstarwr\
can also be retrieved by using the representation in the R.H.S. of \strG\ 
with the help of the formula
\eqn\littletest{
\stbac\({e_\st}^{\int_\rR z\cdot{A}}\)
\({e_\st}^{-\int_\rR r(w)\cdot{\A}}\)\stvac
=e^{-\int_\rL r(z)\cdot w}.
}
It is convenient to introduce the string field coherent state
$|{z}\kket$ and its conjugate such that
\eqn\strcoherent{\eqs{
|z\kket=\({e_\st}^{\int_\rR z\cdot{\A}}\)\stvac,
\quad
\bbra w|=\stbac\({e_\st}^{\int_\rR \b{w}\cdot{A}}\),
}}
satisfying that $A(\s)|z\kket=\theta_\rR(\s)z(\s)|z\kket$,
$\bbra w|\A(\s)=\bbra w|\theta_\rR(\s)\b{w}(\s)$.
Therefore, the coherent state in terms of these states becomes
\eqn\coherentstcoherent{
|z\ket=|z_\rL,z_\rR\ket=|-r(z)\kket\bbra\ \b{z}\,|
=|-r(z_\rL)\kket\bbra\ \b{z}_\rR|,
}
and \littletest\ turns into $\bbra\bz|-r(w)\kket=\exp(-\int_\rR r(w)\cdot 
z)$.

We have so far seen that the open string product can be
rephrased by using the ordinary matrix product.
The `integration' of the open string field theory is defined
by using the identity field $|I\ket$ as
$\int|z_\rL,w_\rR\ket =\bra I|z_\rL,w_\rR\ket$, which is
especially used in defining the action of the string field theory.
In our matrix language, this integration can be seen to correspond to
the trace
\eqn\strtrace{
\Tr\,\Big( |z\kket\bbra w| \Big)\equiv \bbra{w}|{z}\kket
=\bra{I\,}|-r(z),\b{w}\ket.
}
Moreover, the inner product $\bra R(1,2)| |z\ket_1|w\ket_2$
can be given by the trace as
\eqn\strinnerproduct{
\bra R(1,2)| |z\ket_1|w\ket_2=\bra r(z)|w\ket=e^{-r(z)\cdot{w}}
=\Tr\,\Big(\ |-r(z)\kket\bbra \bz|\ |-r(w)\kket\bbra\b{w}|\ \Big).
}

\subsec{The Mode Expansion of the String Field Oscillators}

Since the worldsheet oscillator $\a(\s)$ is expanded as
\eqn\expadphi{
\a(\s)=\sum^{\infty}_{n=0}\a_n\,\varphi_n(\s),
}
we also have the mode expansion of $A(\s)$ and $\A(\s)$ as
\eqn\expAphi{
A(\s)=\sum^{\infty}_{n=0}\a_n\,I\,\theta_\rR(\s)\varphi_n(\s),
\qquad
\A(\s)=-\sum^{\infty}_{n=0}\a_n\,I\,\theta_\rR(\s)\varphi_n(\pi-\s),
}
as seen from \stringosci. In this subsection, we show that
these string oscillators can be expanded as
\eqn\expA{
A(\s)=\sqrt{2}\sum^{\infty}_{n=0} A_n\, \varphi_{2n}(\s),
\qquad
\A(\s)=\sqrt{2}\sum^{\infty}_{n=0} \A_n\, \varphi_{2n}(\s),
}
in terms of $\{\varphi_{2n}(\s)\}_{n=0,1,\cdots}$ instead of
$\{\varphi_{n}(\s)\}_{n=0,1,\cdots}$.
Although $A(\s)$ and $\A(\s)$ are defined on the interval $[\pi/2, \pi]$,
the operator $A(\s)$ can be extended to an operator ${\cal A}$ on the interval 
$[0, \pi]$ such that 
${\cal A}(\s)=\theta_\rR(\s)A(\s)+\theta_\rL(\s)A(\pi-\s)$, and similarly 
for ${\cal A}^{\dag}(\s)$. Because these operators ${\cal A}$ and 
${\cal A}^{\dag}$ satisfy the Neumann boundary condition, they are expanded 
in terms of $\{\varphi_{n}(\s)\}_{n=0,1,\cdots}$. 
However they also satisfy ${\cal A}(\s)={\cal A}(\pi-\s)$ 
and are thus expanded by the even-integer modes
$\{\varphi_{2n}(\s)\}_{n=0,1,\cdots}$. The restriction of ${\cal A}(\s)$
and ${\cal A}^{\dag}(\s)$ onto the interval $[\pi/2, \pi]$ gives \expA.

The mapping of the modes $\a_n\,I$ to the modes $A_n$ and $\A_n$ can be
obtained by using the relations
\eqn\mapatoA{
A_n=\sqrt{2}\int^\pi_{\pi\over2}d\s A(\s)\,\varphi_{2n}(\s),
\qquad
\A_n=\sqrt{2}\int^\pi_{\pi\over2}d\s \A(\s)\,\varphi_{2n}(\s),
}
with the substitution of \expAphi\ and inversely
\eqn\mapAtoa{
\sum^{\infty}_{n=0} \a_n\,I\,\varphi_n(\s)
=\theta_\rR(\s)A(\s)-\theta_\rL(\s)\A(\pi-\s).
}
In fact, we obtain
\eqn\mappingaA{
A_n=\sum_{m=0}^{\infty} R_{n m}\a_m\,I,
\qquad
\A_n=\sum_{m=0}^{\infty} L_{n m}\a_m\,I,
}
where the definition of the matrices $R_{nm}$ and $L_{nm}$ is
\eqn\RLmatrix{
R_{nm}=\sqrt{2}\int^{\pi}_{\pi\over2}d\s\, \varphi_{2n}(\s)\, 
\varphi_m(\s),
\qquad
L_{nm}=-\sqrt{2}\int_{0}^{\pi\over2}d\s\, \varphi_{2n}(\s)\, \varphi_m(\s).
}
See appendix A for the properties of these matrices. 
The inverse mapping turns out to be
\eqn\mappingAa{
\a_m\,I=\sum_{n=0}^{\infty} \[ A_n R_{nm} + \A_n L_{nm} \].
}
Thus, all the modes $A_n$ and $\A_n$ are almost in one-to-one
correspondence with all the modes $\a_n$\foot{In order to prove 
the statement, we would need to estimate the Jacobian associated with 
the change of the variables \mappingaA. 
From the condition ${\cal A}(\s)={\cal A}(\pi-\s)$, we can see that 
${\cal A}(\s)$ satisfies the extra boundary condition 
$\partial_{\s}{\cal A}(\s=\pi/2)=0$. Therefore, the mapping \mappingAa: 
$A_n,\A_n\mapsto\a_n$ probably is not completely surjective. 
This subtlety will be discussed in section 5.}. 

Let us consider the coherent state $|z\ket$
in terms of the modes $\A(\s)$.
Although the function $z(\s)$ in the coherent state $|z\ket$ is defined on
$[0, \pi]$ and has the expansion
$z(\s)=\sum^{\infty}_{n=0} z_n \varphi_n(\s)$, its left and right parts 
$\theta_{\rL,\rR}(\s)z(\s)$ can be expanded as
$\theta_{\rL}(\s)z(\s)=\sqrt{2}\sum^{\infty}_{n=0} z_{\rL,n}\varphi_{2n}(\s)$
and 
$\theta_{\rR}(\s)z(\s)=\sqrt{2}\sum^{\infty}_{n=0}z_{\rR,n}\varphi_{2n}(\s)$, 
in the same way as the operators $A(\s)$, $\A(\s)$ in \expA. 
Therefore, from \strG, we obtain
\eqn\strGmode{
|z_\rL, z_\rR\ket={e_\st}^{-\sum_{n=0}^{\infty}z_{\rL,n}\A_n}
\stvac\stbac{e_\st}^{\sum_{n=0}^{\infty}z_{\rR,n}A_n},
}
and the string field coherent state is thus rewritten as
\eqn\strcoherentmode{
|-r(z)\kket={e_\st}^{-\sum_{n=0}^{\infty}z_{\rL,n}\A_n}\stvac
}
in terms of the modes $\A_n$. Making use of the modes $\A_n$,
we can define states with definite occupation numbers of the modes:
\eqn\Fockstate{
|{\bf n}\kket=
\sqrt{1\over n_0!\, n_1!\cdots}
\(\A_0\)^{n_0}\(\A_1\)^{n_1}\cdots\stvac
}
where the infinitely dimensional vector ${\bf n}$ denotes
$(n_0,n_1,\cdots)$.  In terms of such states $|{\bf n}\kket$, we have
\eqn\coherentoccupation{
|-r(z)\kket=\sum^{\infty}_{n_0,n_1,\cdots=0}(-)^{(\sum^{\infty}_{i=0}{n_i})}
{(z_{\rL,0})^{n_0}(z_{\rL,1})^{n_1}\cdots\over \sqrt{n_0!\, n_1!\cdots}}
|{\bf n}\kket.
}
Since \mapatoA\ and the commutation relation \commAAd\ provide the
commutation relations among the modes $A_n$ and $\A_n$ as
\eqn\commmodeA{
\[A_n, \A_m\]_\st=\d_{n,m}I,
\qquad
\[\A_n, \A_m\]_\st=\[A_n, A_m\]_\st=0,
}
the states $|{\bf n}\kket$ satisfy the orthonormality condition
\eqn\orthonorm{
\bbra{\bf m}|{\bf n}\kket=\d_{m_0,n_0}\d_{m_1,n_1}\cdots
\equiv \d_{{\bf m},{\bf n}}.
}

Since we have the mapping from string fields to matrices,
a generic string field can be written as
\eqn\mapphitonm{\eqalign{
|\Psi\ket&=\int[dz d\b{z}]e^{-z\cdot\b{z}}
|z\ket\bra z|\Psi\ket
\cr
&=\int[dz d\b{z}]e^{-z\cdot\b{z}}
\Psi(z)\,|-r(z)\ket\ket\bra\bra{\bz}|,
}}
where $\Psi(z)=\bra z|\Psi\ket$. In this way,
all string fields can be converted to matrices
on ${\cal H}_{str}$, which is spanned by the basis of the vectors
$\{|{\bf n}\kket\}$. In particular, the identity field $I$
is converted to
\eqn\IstrG{\eqalign{
I&=\int[dz d\b{z}]e^{-z\cdot\b{z}
-\int_{\rL}\b{z}\cdot r(\b{z})}
|-r(z)\ket\ket\bra\bra\b{z}|.
}}

Since $|-r(z)\kket$ and $\bbra z|$ are expanded by
$|{\bf n}\kket$ and $\bbra {\bf m}|$, respectively,
a generic string field $\Psi$ can be written as
\eqn\Psinm{
\Psi=\sum_{{\bf n},{\bf m}}\Psi_{{\bf n},{\bf m}}|{\bf n}\kket
\bbra {\bf m}|.
}
Then for the identity field $I$, since $I\st \Psi=\Psi\st I=\Psi$
up to some potential anomalies \refs{\HorowitzYZ,\RastelliZ,\StromingerBN}, 
we expect that the formula 
\eqn\KOInn{
I=\sum_{{\bf n}}|{\bf n}\kket\bbra {\bf n}|
}
holds. In order to prove \KOInn, we have to precisely estimate 
the measure $[dzd\bz]$ in \IstrG\ under the change of the variables: 
$z_n\to z_{\rL,n}, z_{\rR,n}$, which we can perform in a similar way to 
the mapping \mappingAa. To this end, we need to compute the Jacobian 
associated with the change of the variables, 
but we have not done such a computation. 
Instead, we will hereafter assume \KOInn.


\newsec{Some Applications of the String Field Algebra}

In the last section, we have seen that string fields can be expanded
in terms of the operators $|{\bf n}\kket\bbra{\bf m}|$.
As explained in the lecture \Harvey\ (and related references therein),
in the case of noncommutative (NC) $\R^2$, where we have the creation and
annihilation operators $\a$, $a$ satisfying $\[a, \a\]=1$,
the generating function
\eqn\defGzbz{
G(z,\b{z})=|z\ket\bra z|=e^{za^{\dag}}|0\ket\bra 0|e^{\b{z}a}
=\sum_{n,m=0}^{\infty}{z^n\b{z}^m\o\rt{n!m!}}|n\ket\bra m|
}
of the operators $|n\ket\bra m|$ is a key ingredient to find projection
operators and shift operators, of which the latter give a field theoretical 
analog of the partial isometries $U$ and $V$ mentioned in the introduction.
Therefore, before proceeding to the string field theory, let us briefly 
recall
some facts we shall need later.

The defining property of $G(z,\b{z})$ is
\eqn\Grelation{
G(z,\b{z})\st G(w,\b{w})=e^{\b{z}w}G(z,\b{w}),
}
and we impose on $G(z,\b{z})$ the boundary condition 
\eqn\Gzerozero{
G(0,0)=2e^{-x^2}=|0\ket\bra0|.
}
The generating function $P(u)$ of projection operators
$P_n=|n\ket\bra n|$:
\eqn\TaylorPn{
P(u)=\sum_{n=0}^{\infty}P_nu^n
}
can be given by using $G(z,\b{z})$ as
\eqn\relGtoP{
\int{d^2z\o\pi}e^{-|z|^2}G(u^{\hf}z,u^{\hf}\b{z})=P(u).
}
In particular, since the generating function $P(u)$
satisfies $P(0)=P_0$, $P(1)=I$, 
the completeness condition
\eqn\complete{
\sum_{n=0}^{\infty}P_n = I 
}
can also be expressed by \relGtoP\ at $u=1$; 
\eqn\relGtoP{
\int{d^2z\o\pi}e^{-|z|^2}G(z, \b{z})=I.
}
Moreover, using the function $G(z,\b{z})$, we can construct
the shift operator
\eqn\shiftinG{
S=\sum_{n=0}^{\infty}|n+1\ket\bra n|
=\sum_{n=0}^{\infty}\rt{n!(n+1)!}\oint{dz\o z^{n+2}}\oint{d\b{z}\o \b{z}^{n+1}} 
G(z,\b{z}).
}

\subsec{The Generating Function of the Operators
$|{\bf n}\kket\bbra{\bf m}|$}

We can see from the previous section that the generating function $G(z,w)$ 
of the string field theory is given by
\eqn\Gasmatcoh{
G(z,\b{w})=|z\kket\bbra w|,
}
which can indeed be verified from \littletest\ to satisfy the stringy
extension of the defining property \Grelation:
\eqn\Gstrelation{\eqs{
G(z,\bz)\st G(w,\bar{w})&=e^{\int_\rR\bz\cdot{w}}G(z,\bar{w})
\cr
&=e^{\sum_{n=0}^{\infty}\bz_{\rR,n}\cdot w_{\rR,n}}G(z,\bar{w}),
}}
where 
$\theta_\rR(\s)\bz(\s)=\sqrt{2}\sum_{n=0}^{\infty}\bz_{\rR,n}\varphi_{2n}(\s)$
and similarly for $w(\s)$.
\noindent
\noindent
The equation \mapphitonm\ provides the counterpart of \relGtoP\ for
$u=1$;
\eqn\StrelGtoP{
I=\int[dz d\b{z}]e^{-z\cdot\b{z}
-\int_{\rL}\b{z}\cdot r(\b{z})}G(-r(z),z).
}

\subsec{Projection States and Partial Isometries}
In the open string field theory around the tachyon vacuum \RSZone,
projection operators found in \KP\ have been used to describe
$D$-branes as soliton solutions \RSZtwo.
Before that, we had seen a somehow analogous usage of projection operators
in the noncommutative field theory \refs{\GMS,\DasguptaMR,\HKLM}.
Therefore, it is natural to study projections in the algebra of string 
fields.
The defining property of the projection state $|P\ket$ is
$P\st P=P$; more precisely,
\eqn\projction{
|P\ket_1=\ _3\bra r(P)|\ _2\bra r(P)||v_{3}^m(1,2,3)\ket.
}
In general, for orthogonal projectors $P_n$ satisfying
\eqn\PnPmst{
P_n\st P_m=\cob_{n,m}P_m,\qquad \sum_{n=0}^{\infty}P_n=I,
}
the generating function
\eqn\TaylorPn{
P(u)=\sum_{n=0}^{\infty}P_nu^n
}
can be seen to satisfy the equation $P(u)\st P(v)=P(uv)$, to which
one of the solutions is found to be
\eqn\KWNpu{
P(u)=\exp\lf(-{u\o2}a^{\dag}\cdot r(a^{\dag})\ri)|\Om\ket.
}
Note that $P(u)$ interpolates between the $a$-vacuum  and the identity
\eqn\interKWNpu{
P(0)=|\Om\ket,\quad P(1)=\exp\lf(-\hf a^{\dag}\cdot 
r(a^{\dag})\ri)|\Om\ket=I.
}
The projectors $P_n$ can thus be read as 
\eqn\PnKWN{
P_n={1\o n!}\lf(-\hf a^{\dag}\cdot r(a^{\dag})\ri)^n|\Om\ket.
}
However, unfortunately these projection operators $P_n$ in \PnKWN\
do not have finite norms; $\bra P_n|P_n\ket=\infty$.
Incidentally, $P(u)$ in the usual oscillator formalism 
\refs{\Samuel\GrossJ\CST\ItohWM{--}\OhtaWN} is presented
in appendix B, and its norm also turns out to be ill-defined.

Besides the obvious example of the projectors
$P_{\bf n}=|{\bf n}\kket\bbra{\bf n}|$,
an alternative projector is given by
$P(z)=e^{-\hf z\cdot r(z)}|z\ket$,
where the reality condition $\bra P|=\bra r(P)|$
enforces
$z(\s)=\sum^{\infty}_{n=0}z_n\varphi_n(\s)$ to satisfy
that $\bz_{n}=(-)^{n+1}z_n$.
This projection operator is associated with the
coherent state $|z\ket\ket$ as
\eqn\projtoz{
P(z)={|z\ket\ket\bra\bra z|\o \bra\bra z|z\ket\ket}=
e^{\int_{\rR}-z\b{z}-zr(\a)+\b{z}\a}|\Om\ket.
}

The projection operator $P_{\bf 0}=\stvac\stbac=|\Om\ket$ 
satisfies $a(\s)|\Om\ket=0$, which suggests that $P_{\bf 0}$ 
does not correspond to any extended objects\foot{
The observation in this paragraph is mainly based on work 
in collaboration with Kazuki Ohmori \KOO.}. 
Indeed, $P_{\bf 0}=P(z=0)$ is an instanton-like object, 
since its profile is found to be gaussian 
in all the directions, including the `time' direction. 
Furthermore, the projector $P(z)$ can also be seen to be an instanton. 
Among the parameters $z_n$, the zero mode $z_0$ 
represents the location of the instanton $P(z)$ in spacetime. 
As in \RSZone, if the ghost part of these solutions is universally given, 
the ratio of the norms of these solutions is equal to the ratio of 
their tensions in the tachyon string field theory \RSZone. 
Since the norm of the projector $P(z)$ is the same as that of $P_{\bf 0}$, 
the moduli parameters $z(\s)$ may correspond to collective coordinates 
or the gauge redundancy. We shall leave a closer inspection of this problem 
to the future. The other projections $P_{\bf n}$ are also instanton-like. 
Thus, we may regard the projection operators $\sum^{N}_{k=1}P_{{\bf n}_k}$ 
as $N$-instanton solutions, though we have not fully uncovered the physics of 
such objects. 

Finally, we shall discuss partial isometries in the cubic string field theory 
in our matrix language. As in \refs{\HKL,\Harvey}, 
our experience with noncommutative solitons on NC $\R^2$ shows 
one obvious example of partial isometries
\eqn\shiftAn{
U={1\o\rt{A_n^{\dag}\st A_n+I}}\st A_n,\quad 
V=A_n^{\dag}\st{1\o\rt{A_n^{\dag}\st A_n+I}}. 
}
The commutation relation $[A_n, \A_m]_\st=\d_{n,m}I$ ensures that 
$U\st V=I$, by which we can verify that $V\st U$ is a projector. 
Since $A_n\stvac=0$, we can see that $V\st U\not=I$.  
These operators $U$, $V$ are thus the shift 
operators associated with the $n^{\rm th}$ string field oscillator.

A further example can be given by
\eqn\UVNCRfour{\eqalign{
U&=I-\sum_{n_0=0}^{\infty}|n_0,{\bf 0}\kket\bbra n_0,{\bf 0}|
+ \sum_{n_0=0}^{\infty}|n_0,{\bf 0}\kket\bbra n_0+1,{\bf 0}|, 
\cr
V&=I-\sum_{n_0=0}^{\infty}|n_0,{\bf 0}\kket\bbra n_0,{\bf 0}|
+ \sum_{n_0=0}^{\infty}|n_0+1,{\bf 0}\kket\bbra n_0,{\bf 0}|,
}}
with the defining property 
\eqn\UVfourrel{
U\st V=I,\quad V\st U=I-|\Om_{\rm str}\kket\bbra \Om_{\rm str}|.
}
In this case, the pair $(U, V)$ is an analogue of the one used in 
the discussion of noncommutative $U(1)$ instantons on $\R^4$\Furuuchi.

Before concluding this section, we should point out that, 
in order for these partial isometries to give classical solutions, 
The action of the BRS charge $Q$ on the resulting projector $V\st{U}$ 
must vanish, as explained in the introduction. 
For the BRS charge $Q$ in a usual flat Minkowski 
spacetime, however, this turns out not to be the case.

\newsec{Conclusion and Discussion}

In this paper, we have introduced a new presentation of the string vertices 
in the cubic string field theory. We have seen that the new string 
vertices have very simple structures and facilitate some computations  
which have been very complicated. 
By making use of these new vertices, we have tried to 
reformulate the open string field product or the $\st$-product as 
the usual matrix multiplication\foot{In the original paper \Witten, 
Witten has already discussed such a possibility (see also 
\refs{\ChanIE,\BordesXH,\AbdurrahmanGU} and for very recent discussions, 
\refs{\RSZthree,\GT}). }.
Indeed, we believe that to some extent in this paper, we have been 
successful in obtaining huge-sized matrices as open string fields. 
However, we have also been cavalier on two points: our treatment of the 
midpoint and some anomalies \refs{\HorowitzYZ,\RastelliZ,\StromingerBN} 
associated with singularities of the $\st$-product.

Since our representation of the string vertices satisfies the overlap 
condition even at the midpoint, we believe there is nothing wrong 
with our string vertices. However, our discussion in section 3 about splitting 
the operator $\a(\s)\,I$ into $A_n$ and $\A_n$ is subtle, because of 
the usage of the functions $\theta_{\rL,\rR}(\s)$ as projectors; 
$\theta_{\rL,\rR}(\s)^2=\theta_{\rL,\rR}(\s)$. In addition, 
our Fourier transform $A_n$, $\A_n$ of the left and right halves of 
open strings does not seem enough to describe all the degrees of freedom 
of open strings, as mentioned in the footnote of section 3. 
Although, in the original suggestion \Witten, one additional degree of 
freedom associated with the midpoint of strings is added, it is 
not clear to us whether we should also introduce such a degree of freedom. 

The string oscillators are defined by using the identity state $I$. 
It is known \refs{\StromingerBN,\HorowitzYZ} that states like $Q_\rL\,I$ 
in \HorowitzLRS\ have anomalies in calculations with the string vertices. 
Since the modes $A_n$ and $\A_n$ are also defined by acting on $I$ with 
the integration of $\a(\s)$ over half of the open strings,  
there might exist some potential anomalies in the matrix multiplication 
in our formulation. 

In this paper, we have restricted our attention to only the matter part of 
open strings. As for the ghost part, we have not been successful in 
finding a simple representation of the string vertices by using 
the fermionic ghosts $b(\s)$ and $c(\s)$. 
However, the ghost part of the vertices might be able to be simplified 
by using the bosonized ghost in a similar way to the matter part of our 
vertices. Since the construction of the ghost part requires the insertion of 
the ghosts at the midpoint, we would have more serious problems 
concerning the midpoint, as we have discussed in the above paragraphs. 

We could have obtained our matrix formulation of the cubic open string 
field theory in the background $B$-field \refs{\KT,\Sugino}. 
It would be interesting to explore what is happening in our matrix 
formulation in the large $B$ limit and to compare that with the discussion 
in \refs{\WittenNCG,\Snbl}. 

As in the paper \RSZtwo, it is found that the projectors of the matter part 
give the ratios between the $D$-brane solutions in the tachyon string field 
theory \RSZone\ with the ansatz of the universality of the ghost parts by
making use of the method in \KP. It is possible to apply an analogous method 
to our vertices, and to find the projectors which are supposed to be $D$-branes. 
However, we can analytically calculate the norm of those solutions and 
find no finite tensions of those solutions except the instanton-like solutions 
given in section 4. By the tensions we mean the norms of the solution 
divided by the volume factor of the corresponding $D$-brane worldvolume\foot{
This is also based on \KOO, as mentioned in section 4.}. 
So far, we have not succeeded in finding any brane solutions with finite norms 
by using our three-string vertex. Since it is likely that the norm of 
such $D$-brane solutions can be computed analytically by using our vertex, 
the discovery of the solutions gives a nontrivial check on the interpretation 
of them as $D$-branes. 

As discussed in \WittenDK, the charges of $D$-branes are classified by
K-theory and it is closely related to the topological
configurations of tachyon fields.
Although $D$-branes in bosonic string theory do not carry any
charges, it may be interesting to study K-theory of our matrix algebra 
\WittenK\ of string fields.


\vskip 0.5in
\noindent
{\sl Note added}: after completing this work, we received
the papers \refs{\RSZthree,\GT} concerning a similar idea
to ours: splitting open strings into their left and right halves.
However, on the whole, our approach is slightly different from theirs 
\refs{\RSZthree,\GT}, especially on the introduction of the new 
representation of the string vertices. As a technical point, 
upon splitting open strings, we used a different Fourier expansion of
the `left' and `right' halves from theirs \refs{\RSZthree,\GT}. 
It is at present unclear to us how these expansions are precisely 
related to each other. 

Furthermore, from \refs{\RSZthree,\GT} 
we learnt of the references \refs{\ChanIE,\BordesXH,\AbdurrahmanGU}, 
where earlier discussions (also even earlier references therein) 
about the idea of splitting open strings 
are available, besides the original discussion in \Witten. 

After submitting the original version of this paper 
to the arXiv.org e-print archive, 
we were informed of a paper \ArefevaEG\ arguing another approach 
to formulate string fields as matrices. 

\vskip 2cm
\centerline{{\bf Acknowledgements}}

The authors would like to thank Kazuki Ohmori for his collaboration at 
the early stage of this work and also for his careful reading of the
manuscript and helpful suggestions on it. 
T.$\,$K. is deeply grateful to Taichiro Kugo and Hiroyuki Hata for 
helpful discussions.
K.$\,$O. would like to thank Kazuyuki Furuuchi, Nobuyuki Ishibashi and
Soo-Jong Rey for discussions, and to acknowledge the hospitality of 
theory group at Seoul National University in Korea. 
T.$\,$K. was supported in part by a Grant-in-Aid for Scientific Research
in a Priority Area: ``Supersymmetry and Unified Theory of Elementary
Particles''(\#707), from the Ministry of Education, Culture, Sports,
Science, and Technology.
K.$\,$O. was supported in part by JSPS Research Fellowships for Young
Scientists.

\appendix{A}{Properties of $L_{nm}$ and $R_{nm}$}
To see the properties of the matrices $L_{nm}$ and $R_{nm}$ in \RLmatrix,
it is convenient to introduce other matrices $P_{\rL}$ and $P_{\rR}$
defined by
\eqn\defPLR{
(P_{\rL})_{nm}=\int_{\rL}\vphi_n\vphi_m,\quad 
(P_{\rR})_{nm}=\int_{\rR}\vphi_n\vphi_m.
}
Since $\vphi_n$ transforms under the reflection as
$r(\vphi_n)=(-1)^n\vphi_n$, 
$P_{\rL}$ and $P_{\rR}$ are related by
\eqn\PLRref{
P_{\rL}=CP_{\rR}C,
}
where $C_{nm}=(-1)^n\cob_{nm}$. 
By making use of the completeness condition \intpiphi, \phitocob\ 
of $\vphi_n$,
we can see that $P_{\rL}$ and $P_{\rR}$ are orthogonal projections:
\eqn\PLRproj{
P_{\rL}+P_{\rR}=1,\quad P_{\rL,\rR}^2=P_{\rL,\rR},\quad 
P_{\rL}P_{\rR}=P_{\rR}P_{\rL}=0.
}
These relations can also be understood by noticing that
$P_{\rL,\rR}$ are the representation matrices of 
$\th_{\rL,\rR}(\s)$ on the basis $\{\vphi_n(\s)\}$
\eqn\multhLRrep{
\th_{\rL,\rR}(\s)\vphi_n(\s)=\sum_{m=0}^{\infty}(P_{\rL,\rR})_{nm}\vphi_m(\s).
}

From the definition of $R,L$ in \RLmatrix,
they are written in terms of $P_{\rR,\rL}$ as
\eqn\RLinPLR{
R_{nm}={1\o\rt{2}}\Big[(1+C)P_{\rR}\Big]_{2n,m},\quad
L_{nm}=-{1\o\rt{2}}\Big[(1+C)P_{\rL}\Big]_{2n,m}.
}
Using the relations  \PLRref\ and \PLRproj,
we obtain
\eqn\RLsquare{\eqalign{
\sum_{k=0}^{\infty}R_{kn}R_{km}
&=\hf\Big[P_{\rR}(1+C)^2P_{\rR}\Big]_{nm}=(P_{\rR})_{nm},\cr
\sum_{k=0}^{\infty}L_{kn}L_{km}
&=\hf\Big[P_{\rL}(1+C)^2P_{\rL}\Big]_{nm}=(P_{\rL})_{nm}.
}}
Therefore, $R$ and $L$ satisfy
\eqn\RLsumcob{
\sum_{k=0}^{\infty}(R_{kn}R_{km}+L_{kn}L_{km})=\cob_{nm}.
}
This formula is used to show the relation \mappingAa\ between
$\a_n$ and $A_n,A^{\dag}_n$.
From the relation
\eqn\vphitwon{
\int_{\rR}\vphi_{2n}\vphi_{2m}=\int_{\rL}\vphi_{2n}\vphi_{2m}=\hf\cob_{nm},
}
we can also see that
\eqn\LRrsum{
\sum_{k=0}^{\infty}R_{nk}R_{mk}=\sum_{k=0}^{\infty}L_{nk}L_{mk}=\cob_{nm},
\quad \sum_{k=0}^{\infty}R_{nk}L_{mk}=0.
}

\appendix{B}{Generating Function of Projections using
Gross-Jevicki Vertex}
In this appendix, we consider the generating function of
orthogonal projections using the Gross-Jevicki vertex.
We follow the notation in \RSZtwo.
The function $P(u)$ satisfying $P(u)\st P(v)=P(uv)$ is found to be
\eqn\PuinSFTbose{
P(u)=\det{}^{\hf}{(1-X)(1+T)\o Tu+1}\exp\lf(-\hf a^{\dag}
CT(u)a^{\dag}\ri)|0\ket,
}
where
\eqn\defTu{
T(u)={u+T\o Tu+1}.
}
This can be checked by using the formula
\eqn\TonestTtwo{
e^{-\hf a^{\dag}CT_1a^{\dag}}|0\ket\st e^{-\hf a^{\dag}CT_2a^{\dag}}|0\ket
=\det{}^{-\hf}[1-X(T_1+T_2)+XT_1T_2]e^{-\hf a^{\dag}CT_{12}a^{\dag}}|0\ket,
}
where
\eqn\Tonetwodef{
T_{12}={X-X(T_1+T_2)+T_1T_2\o 1-X(T_1+T_2)+XT_1T_2}.
}
Here we assumed that $C,T_1,T_2$ and $X$ commute with each other.
Note that $P(u)$ interpolates between the matter part of ``sliver''
\refs{\RastelliZ,\KP,\RSZtwo} and the identity:
\eqn\Pzeroone{
P(0)=|\infty\ket_m,\quad P(1)=I.
}

\listrefs

\end